\documentclass{PoS}
\pdfoutput=1

\title{The High Energy Radiation Pattern from {\tt BFKLex}}

\ShortTitle{The High Energy Radiation Pattern}

\author{\speaker{G. Chachamis}\\
        Instituto de F{\' \i}sica Te{\' o}rica UAM/CSIC, Nicol{\'a}s Cabrera 15\\
        \& Universidad Aut{\' o}noma de Madrid, E-28049 Madrid, Spain.
        E-mail: \email{chachamis@gmail.com}}

\author{A. Sabio Vera\\
 Instituto de F{\' \i}sica Te{\' o}rica UAM/CSIC, Nicol{\'a}s Cabrera 15\\
        \& Universidad Aut{\' o}noma de Madrid, E-28049 Madrid, Spain.
        E-mail: \email{a.sabio.vera@gmail.com}}

\abstract{We discuss a recent study on high-energy jet production in the multi-Regge limit done with the use of the Monte Carlo event generator {\tt BFKLex} which includes collinear improvements in the form of double-log contributions. We will show results for the average transverse momentum and azimuthal angle of the final state jets when at least one of them is very forward in rapidity and another one is very backward. We also discuss the introduction of a new observable which accounts for the average rapidity ratio among subsequent emissions.}

\FullConference{XXIV International Workshop on Deep-Inelastic Scattering and Related Subjects\\
		11-15 April, 2016\\
		DESY Hamburg, Germany}

\begin{document}

\section{Introduction}

The applicability of the Balitsky-Fadin-Kuraev-Lipatov (BFKL) framework in 
multi-jet production at the LHC is a topical field of study. 
The important question is whether
BFKL dynamics can affect observables at hadronic colliders  at current LHC energies or, in other words,
whether pre-asymptotic BFKL effects are important.
Here we will indeed discuss how pre-asymptotic effects do show up at present energies assuming BFKL evolution at
leading order (LO)~\cite{Lipatov:1985uk,Balitsky:1978ic,Kuraev:1977fs,Kuraev:1976ge,Lipatov:1976zz,Fadin:1975cb} and next-to-leading order (NLO)~\cite{Fadin:1998py,Ciafaloni:1998gs} with collinear improvements.

In the following we will discuss events with one forward jet (rapidity $y_a$)
and one backward jet  (rapidity $y_b$) such that $Y=y_a-y_b$ is large. Then the cross section
can be written as
\begin{eqnarray}
\sigma (Q_1,Q_2,Y) = \int d^2 \vec{k}_A d^2 \vec{k}_B \, {\Phi_A(Q_1,\vec{k}_a) \, 
\Phi_B(Q_2,\vec{k}_b)} \, {f (\vec{k}_a,\vec{k}_b,Y)}\,
\end{eqnarray}
where, $\Phi_{A,B}$ are the impact factors and $f$ is the BFKL gluon Green function.

We work within the NLO approximation and thus, a dependence on the 
renormalisation scale  and the energy scale cannot be avoided~\cite{Forshaw:2000hv,Chachamis:2004ab,Forshaw:1999xm,Schmidt:1999mz}. One can write the gluon Green function in an iterative way
as a sum of phase-space integrals in rapidity and transverse momentum both at LO~\cite{Schmidt:1996fg} and NLO~ \cite{Andersen:2003an,Andersen:2003wy}. It reads
\begin{eqnarray}
f &=& e^{\omega \left(\vec{k}_A\right) Y}  \Bigg\{\delta^{(2)} \left(\vec{k}_A-\vec{k}_B\right) + \sum_{n=1}^\infty \prod_{i=1}^n \frac{\alpha_s N_c}{\pi}  \int d^2 \vec{k}_i  
\frac{\theta\left(k_i^2-\lambda^2\right)}{\pi k_i^2} \nonumber\\
&&\hspace{-1.2cm}\int_0^{y_{i-1}} \hspace{-.3cm}d y_i e^{\left(\omega \left(\vec{k}_A+\sum_{l=1}^i \vec{k}_l\right) -\omega \left(\vec{k}_A+\sum_{l=1}^{i-1} \vec{k}_l\right)\right) y_i} \delta^{(2)} \hspace{-.16cm}
\left(\vec{k}_A+ \sum_{l=1}^n \vec{k}_l - \vec{k}_B\right) \hspace{-.2cm}\Bigg\}, 
 \end{eqnarray}
where 
$
\omega \left(\vec{q}\right) = - \frac{\alpha_s N_c}{\pi} \log{\frac{q^2}{\lambda^2}} 
$
is the gluon Regge trajectory which depends on an infrared divergencies regulator $\lambda$. 
This iterative solution is implemented in the Monte Carlo code {\tt BFKLex}
which has already been used for various applications~\cite{Chachamis:2013rca,Caporale:2013bva,Chachamis:2012qw,Chachamis:2012fk,Chachamis:2011nz,Chachamis:2011rw}. 

The BFKL approach can be affected in the collinear regions of the phase space by a double-log term in the NLO BFKL kernel which has to be resummed to all orders. Only then one can hope for a stable behaviour  
of the BFKL cross sections~\cite{Salam:1998tj,Ciafaloni:2003ek}.
In Ref.~\cite{Vera:2005jt}, the collinear corrections were resummed to all-orders using the substitution
\begin{eqnarray}
\theta \left(k_i^2-\lambda^2\right) \to \theta \left(k_i^2-\lambda^2\right)  + \sum_{n=1}^\infty 
\frac{\left(-\bar{\alpha}_s\right)^n}{2^n n! (n+1)!} \ln^{2n}{\left(\frac{\vec{k}_A^2}{\left(\vec{k}_A+\vec{k}_i\right)^2}\right)}. 
\label{SumBessel}
\end{eqnarray}
Moreover, in~\cite{Vera:2005jt} one can see that this expression resums to a Bessel function of the first kind 
(see also~\cite{Iancu:2015vea}). 
Application of the above resummation in various phenomenological works
(not based on Monte Carlo techniques) show agreement with experimental analyses~\cite{Vera:2006un,Vera:2007kn,Caporale:2007vs,Vera:2007dr,Caporale:2008fj,Hentschinski:2012kr,Hentschinski:2013id,Angioni:2011wj,Caporale:2013uva,Chachamis:2015ona}.

Recently~\cite{Chachamis:2015zzp}, we implemented the procedure described in Eq.~(\ref{SumBessel}) in  the {\tt BFKLex} Monte Carlo event generator and studied its effect in the behavior of the gluon Green function. Here we discuss a study done on the radiation pattern of the final states in exclusive production of jets within the BFKL framework~\cite{Chachamis:2015ico}.

\section{Averages of characteristic quantities in multi-jet events}

We are interested in configurations with $N+2$ jets, out of
which one jet with transverse momentum $k_a$  is very forward
and another one with momentum $k_b$ is very backward such that the rapidity distance between the two jets,
$y_a-y_b$, is large, analogous to the Mueller-Navelet~\cite{Mueller:1986ey} jets.
Therefore, a typical event is characterised
by $k_a$, $k_b$ and a number $N$ of further final-state jets for which we define three variables: the modulus of the transverse momentum, $|k_i|$,
the azimuthal angle $\phi_i$ and the rapidity $y_i$, with $1 \le i \le N$. 

We may now introduce three distinct averages for the jets in each event: 
the average of the modulus of their transverse momentum ($\langle p_t \rangle$), of their azimuthal angle ($\langle \phi \rangle$) and of the rapidity ratio ($\langle {\mathcal R}_y \rangle$) between subsequent jets:
\begin{eqnarray}
\langle p_t \rangle &=& \frac{1}{N} \sum_{i=1}^{N} |k_i|;
\label{eq:observable1}\\
\langle \phi \rangle &=& \frac{1}{N} \sum_{i=1}^{N} \phi_i;
\label{eq:observable2}\\
\langle {\mathcal R}_y \rangle &=& \frac{1}{N+1}  \sum_{i=1}^{N+1} \frac{y_i}{y_{i-1}}, \,\,\,
\mathrm{with}\,\,\, y_0 = y_a, y_{N+1} = y_b = 0 \,\,\, \mathrm{and} \,\,\, y_{i-1} > y_i.
\label{eq:observable3}
\end{eqnarray}

We consider two different configurations for the transverse momenta
of the forward/backward jets:
i) $k_a = 10$ GeV, $k_b = 12$ GeV,
ii) $k_a = 10$ GeV, $k_b = 20$ GeV and three different rapidity
differences $y_a-y_b = 4, 6, 8$. For each of these cases we have run {\tt BFKLex} 
in order to produce differential distributions for the observables 
in Eqs.~(\ref{eq:observable1}),~(\ref{eq:observable2}),
and~(\ref{eq:observable3}) at LO and NLO+Double Logs also
using the anti-$k_t$ jet algorithm~\cite{Cacciari:2008gp}
in the {\tt FastJet} implementation~\cite{Cacciari:2011ma,hep-ph/0512210}
with a jet radius of $R=0.7$. 

\begin{figure}
\begin{center}
\hspace{-1.05cm}
\includegraphics[height=6.2cm]{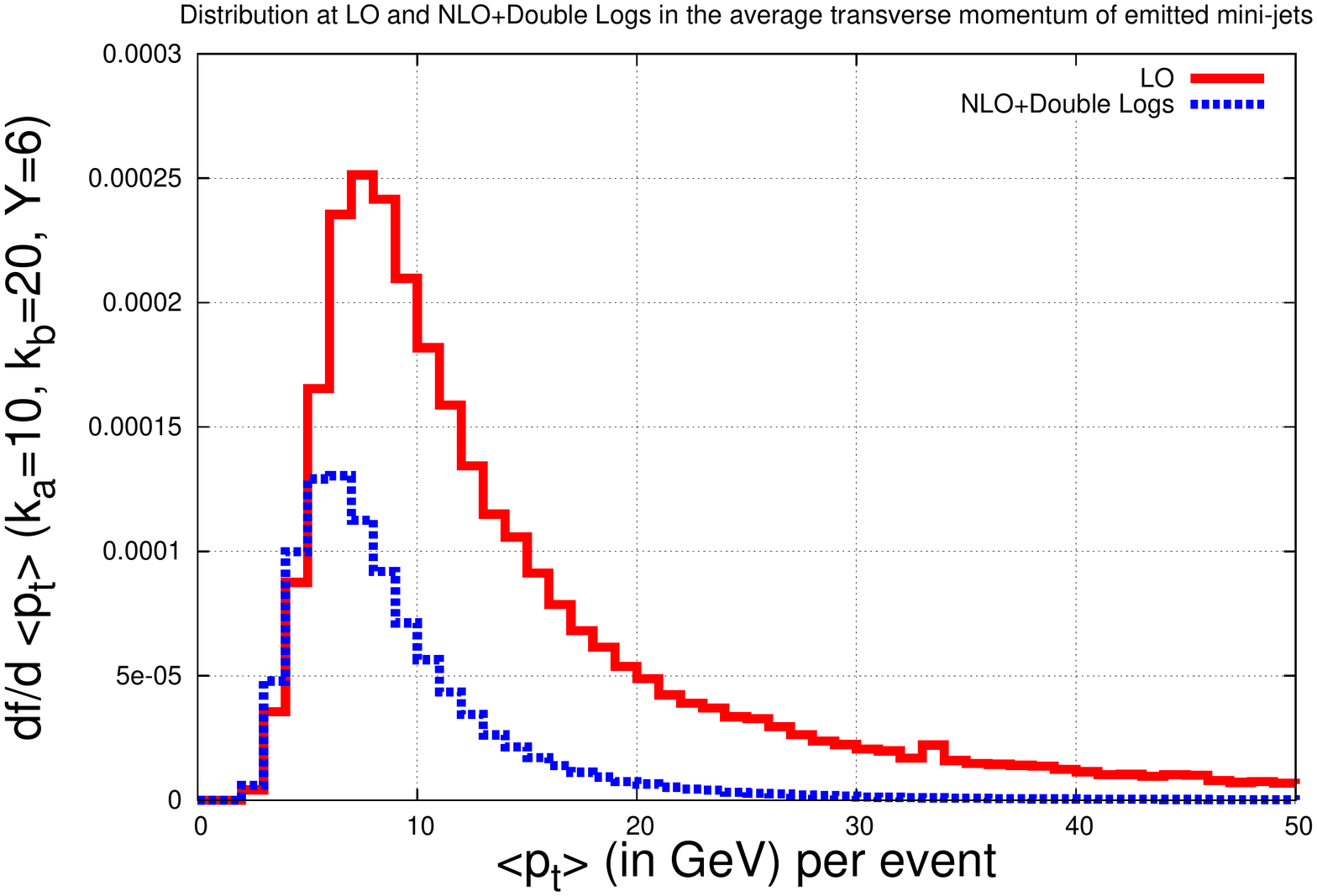}\includegraphics[height=6.2cm]{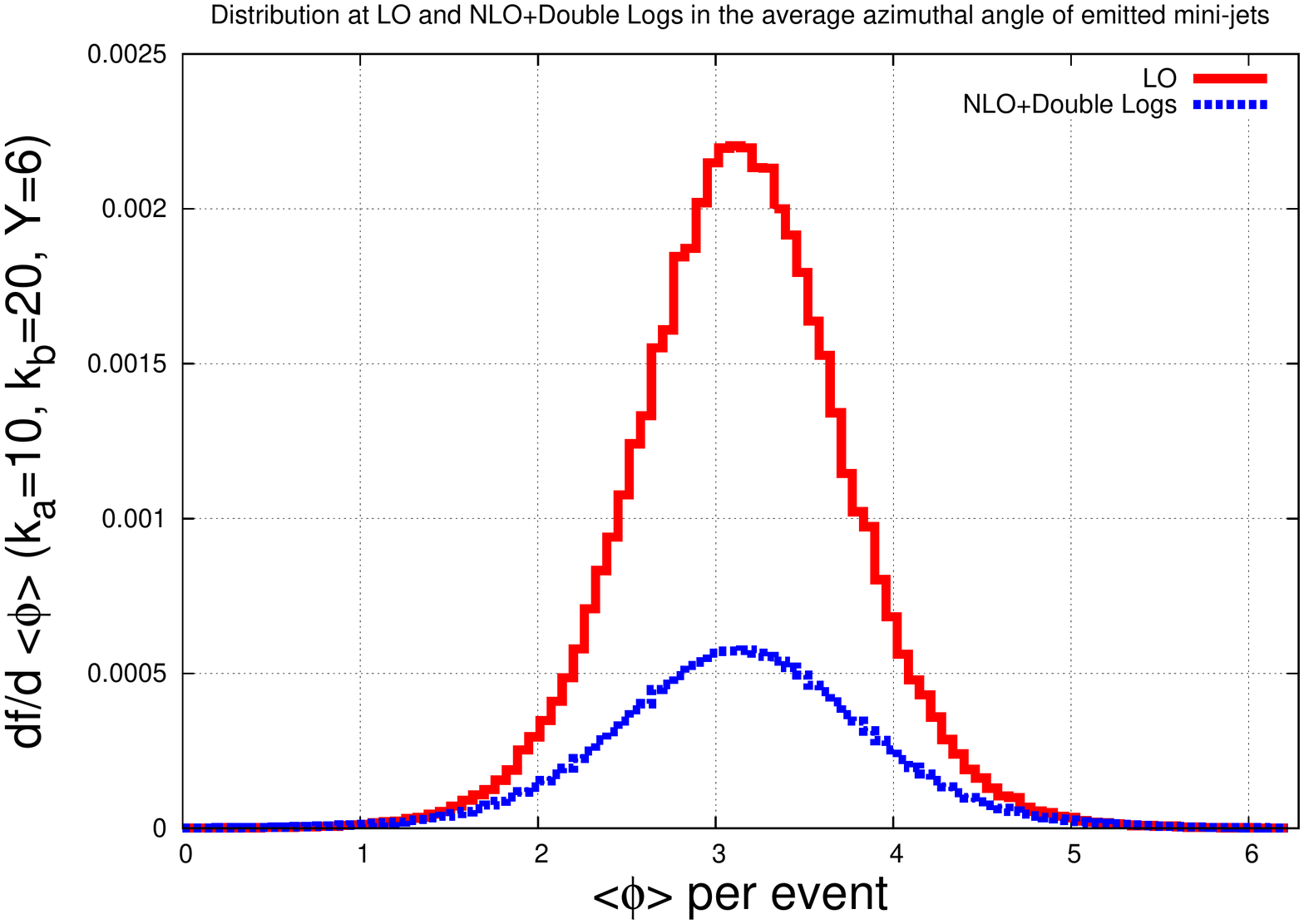}\\
\vspace{-.5cm}
\hspace{-1.05cm}
\includegraphics[height=6.2cm]{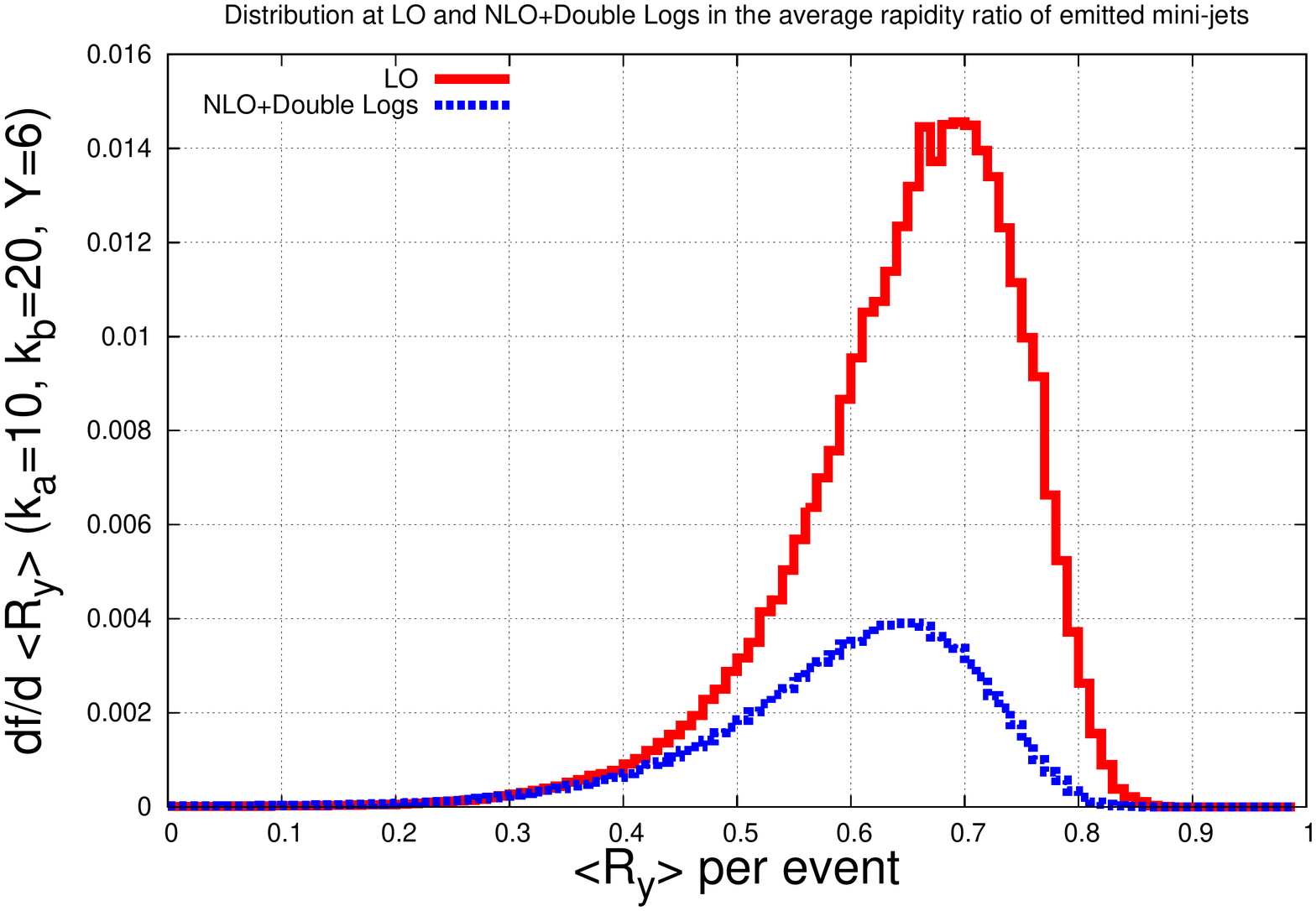}
\end{center}
\vspace{-1cm}
\caption{Average transverse momentum of emitted mini-jets per event (top left), 
average azimuthal angle of emitted mini-jets per event (top right) and 
average rapidity ratios of emitted mini-jets per event (bottom), all at $Y=6$.}
\label{Plots}
\end{figure}

The results for $Y=6$ and $k_a = 10$ GeV, $k_b = 20$ GeV at LO and NLO+Double Logs are seen 
in  Fig.~\ref{Plots}. 
The average $p_t$ (top left) is smaller when running
the  NLO kernel together with double logarithmic collinear terms than the LO one. 
 A sizeable contribution to the Green function and, consequently, 
 to the cross section comes from jets  with a large transverse momentum 
 that cannot be described as mini-jets. 
 This characteristic is more
 pronounced at LO than at NLO and beyond and this change
 originates at the diminution of the diffusion picture at LO+Double Logs~\cite{Chachamis:2015zzp}. Obviously,
 the areas under the differential distributions are smaller 
 when we consider NLO+Double Logs runs for any $Y$ 
 due to the Pomeron intercept getting much smaller when going beyond LO. 

With regard to the azimuthal angle average, Fig.~\ref{Plots} top right, 
we note that at LO the largest part of the final state jets
carries an average angle in between $ \simeq \pi \pm 1$. This trend does not change when $Y$ varies. Actually,
the picture we have just drawn  doesn't seem to change much when we consider the NLO+Double Logs case. The whole description here though is very crude to the point it may be deceptive. We need to remember that the azimuthal decorrelation of the two utmost jets is going to be highly dependant on the exact form of the azimuthal 
angle average distribution. Therefore, it would be very interesting to study with {\tt BFKLex} generalised ratios 
of correlation functions of products of cosines of azimuthal angle differences among the tagged jets
like the ones defined in~\cite{Caporale:2015vya,Caporale:2016soq,Caporale:2015int,Caporale:2016xku}.

Turning the discussion toward the mean distance in rapidity between the different final state jets 
we should point out that the distributions for these ratios have their maximum at
 $\langle {\mathcal R}_y \rangle$ larger than 0.5. 
We also observe that these differential distributions are generally broad, which leads us to the conclusion 
that preasymptotic configurations away from multi-Regge kinematics contribute significantly 
to the cross section. 

We consider the observables discussed here worthy of dedicated
experimental analyses at the LHC. It will be decisive to 
verify if any pre-asymptotic effects can be seen in the data. We think that
the featured broadening of the differential distributions 
for $\langle p_t \rangle$, $\langle \phi \rangle$ and $\langle {\mathcal R}_y \rangle$ is a distinct signal of BFKL 
dynamics and needs to be further investigated both theoretically and experimentally. 

\section{Summary \& Outlook}

We defined a new set of three observables in multi-jet final states which can be of great importance
in the effort of finding distinct BFKL signals at the LHC. 
 The average transverse momentum,  the average azimuthal angle and the average ratio of jet rapidities
 are actually probing the ``anatomy" of the BFKL ladder. 
 We used the Monte Carlo event generator {\tt BFKLex} to produce differential distributions of the three
 observables  after implementing the collinear resummation of the NLO kernel. In order  to
 perform our numerical study within collinear factorization we demand configurations
similar to the usual Mueller-Navelet jets (one very forward and one very backward jet of similar size)
but with additional hard jets  in the final state.
It is compulsory to study observables of very exclusive character after restrictive kinematical cuts
are imposed in order to find a reliable window of applicability of the BFKL formalism. Nevertheless, once
we know exact regions of the phase-space where the BFKL dynamics is dominant, we will be able 
to extend the formalism to other less favourable experimental configurations.

\begin{flushleft}
{\bf \large Acknowledgements}
\end{flushleft}
GC acknowledges support from the MICINN, Spain, under contract FPA2013- 44773-P, ASV thanks the Spanish Government (MICINN (FPA2015-65480- P)) and both to the Spanish MINECO Centro de Excelencia Severo Ochoa Programme (SEV-2012-0249) for support.

\end{document}